\documentclass[12pt,preprint]{aastex}
\def\ltsima{$\; \buildrel < \over \sim \;$}
\def\gtsima{$\; \buildrel > \over \sim \;$}
\def\lsim{\lower.5ex\hbox{\ltsima}}
\def\gsim{\lower.5ex\hbox{\gtsima}}
\def\lapp{\ifmmode\stackrel{<}{_{\sim}}\else$\stackrel{<}{_{\sim}}$\fi}
\def\gapp{\ifmmode\stackrel{>}{_{\sim}}\else$\stackrel{<}{_{\sim}}$\fi}

\newdimen\minuswidth    
\setbox0=\hbox{$-$}
\minuswidth=\wd0
\catcode`@=\active
\def@{\kern\minuswidth}
\newdimen\digitwidth    
\setbox0=\hbox{\rm0}
\digitwidth=\wd0
\catcode`!=\active
\def!{\kern\digitwidth}
 
\shorttitle{Star counts in NGC6266} 
\shortauthors{Beccari et al.} 
 
\begin{document} 
 
\title{The dynamical state and blue straggler population of the
  globular cluster NGC~6266 (M62)
\footnote{Based on observations with the NASA/ESA HST, obtained at the
Space Telescope Science Institute, which is operated by AURA, Inc.,
under NASA contract NAS5-26555. Also based on WFI observations
collected at the European Southern Observatory (ESO), La Silla, Chile,
within the observing programmes 62.L-0354 and 64.L-0439.} }

\author{G. Beccari\altaffilmark{1,2,3},
F.R. Ferraro\altaffilmark{4},
A. Possenti\altaffilmark{5},
E. Valenti\altaffilmark{4,1},
L. Origlia\altaffilmark{1}, 
R.T. Rood\altaffilmark{6}}
\affil{\altaffilmark{1} INAF--Osservatorio Astronomico 
di Bologna, via Ranzani 1, I--40127 Bologna, Italy, 
giacomo.beccari@bo.astro.it}
\affil{\altaffilmark{2} Dipartimento di Scienze della Comunicazione, 
Universit\`a degli Studi di Teramo, Italy}
\affil{\altaffilmark{3} INAF--Osservatorio Astronomico 
di Collurania, Via Mentore Maggini, I--64100 Teramo, Italy}
\affil{\altaffilmark{4} Dipartimento di Astronomia, Universit\`a 
degli Studi di Bologna, via Ranzani 1, I--40127 Bologna, Italy, 
francesco.ferraro3@unibo.it}
\affil{\altaffilmark{5} INAF--Osservatorio Astronomico di Cagliari,
Loc. Poggio dei Pini, Strada 54, I--09012 Capoterra, Italy,
possenti@ca.astro.it}
\affil{\altaffilmark{6} Astronomy Department, University of Virginia, 
Charlottesville, VA, 22903, rtr@virginia.edu}
\medskip

\begin{abstract}
We have used a proper combination of multiband high-resolution {\it
HST-WFPC2} and wide-field ground based observations to image the
galactic globular cluster NGC~6266 (M62).  The extensive photometric data
set allows us to determine the center of gravity and to construct the
most extended radial profile ever published for this cluster
including, for the first time, detailed star counts in the very inner
region.  The star density profile is well reproduced by a standard
King model with an extended core ($\sim 19''$) and a modest value of
the concentration parameter ($c=1.5$), indicating that the cluster has
not-yet experienced core collapse.

The millisecond pulsar population (whose members are all in binary
systems) and the X-ray emitting population (more than 50 sources
within the cluster half mass radius) suggest that NGC~6266 is in a
dynamical phase particularly active in generating binaries through
dynamical encounters. UV observations of the central region have been
used to probe the population of blue straggler stars, whose origin
might be also affected by dynamical interactions. The comparison with
other globular clusters observed with a similar strategy shows that
the blue straggler content in NGC~6266 is relatively low,
suggesting that the formation channel that produces binary systems
hosting neutron stars or white dwarfs is not effective in
significantly increasing the blue straggler population. Moreover, an
anticorrelation between millisecond pulsar content and blue straggler
specific frequency in globular cluster seems emerging with increasing
evidence.
\end{abstract}


\section{Introduction} 
\label{intro}
In the last decades it has become obvious that some stars observed in
Globular Clusters (GCs) are not the products of the passive evolution
of isolated stars. Stellar encounters can produce binary systems and
affect the evolution of the new produced binaries as well as any
binaries that might have existed since the formation of the
cluster. Thus the stellar content of a GC is intimately linked with
the dynamical history of the cluster. In fact, the Color-Magnitude
Diagrams (CMDs) of the core regions of GCs recently observed using
HST-WFPC2 have shown the presence of a variety of exotic stellar
objects, whose formation and evolution may be strongly affected by
dynamical interactions. These include Low-Mass X-ray
Binaries (LMXBs), (e.g. Heinke et al. 2001,
Edmonds et al. 2003), Cataclysmic Variables (CVs) (e.g. Cool et
al. 1998), and Blue Straggler Stars (BSSs) (e.g. the cases of
47 Tuc, Ferraro et al 2004; M3, Ferraro et al. 1997; M80,
Ferraro et al., 1999a; M30, Guhathakurta et al., 1998, Ferraro et
al. 2003a, hereafter F03).

BSS were first discovered by Sandage (1953) in the globular cluster
M3, and manifest themselves in the CMD as an extension of the Main
Sequence (MS) above the Turn-Off (TO), mimicking MS stars with 
larger initial masses. There are two viable proposed mechanisms for BSS
generation: the first is mass exchange in a primordial binary system
and the second is the merger of two stars induced by stellar
interactions (either single or binary) in a dense stellar environment.
The fact that BSSs have been found in all properly observed stellar systems
make them a powerful tracer of dynamical history of the parent
cluster.

Millisecond Pulsars (MSPs) may also be the by-product of collisions in
the dense environment of a cluster. They form in binaries containing a
neutron star (NS) which is eventually spun up through mass accretion
from the evolving companion. Since they are point-like objects and (in
many cases) extremely stable clocks, they are invaluable tools to
investigate the binary evolution in a very dense stellar environment
(e.g.  Rappaport et al. 2001). 
It is important to study the overall
properties of GCs hosting MSPs and to compare the distributions of the
objects which may be significantly affected by
dynamical encounters. Hence we have undertaken a long-term project
to observe in the optical band the GCs which are rich in known
MSPs, with particular emphasis on the BSS population. After having
examined 47~Tucanae (Ferraro et al. 2001, 2004; Mapelli et al. 2004)
and NGC~6752 (Ferraro et al. 2003b, Sabbi et al.  2004), we now turn
our attention to NGC~6266.

NGC~6266 (M62) is a high density ($\log \rho_c \sim$ 5.34), highly
reddened ($E(B-V)\sim$ 0.47) GC (Jacoby 2002; Possenti et al. 2003,
hereafter P03) and is one of the most massive ($M_v=-9.19$, Harris
1996{\footnote{For all references to Harris (1996) we have used the
updated data set at the web site
http://www.physics.mcmaster.ca/Globular.html}}).  It has been
classified as a possible Post Core Collapse (PCC) GC in the
compilation by Djorgovski \& Meylan (1993, hereafter DM93). Six binary
MSPs have been recently discovered in this cluster (D'Amico et
al. 2001a, 2001b; Jacoby et al. 2002; P03). NGC~6266 ranks fourth of
the GCs in wealth of MSPs, after Terzan~5, 47~Tucanae and M15.
Surprisingly, all MSPs in NGC~6266 are in binary systems (P03). As
discussed in P03, the absence of known isolated MSPs in NGC~6266
cannot simply be ascribed to a selection effect since, for a given
spin period and flux density, an isolated MSP is easier to detect than
a binary MSP. If such a lack of isolated pulsars is not a statistical
fluctuation, it must be somehow related to their formation mechanism
and to the dynamical state of the cluster.  Moreover, recent deep {\it
Chandra} X-ray observations of NGC~6266 have revealed a very large
number of X-ray sources---51 sources were detected within the cluster
1\farcm 23 half mass radius (Pooley et al. 2003), indicating that an
high number of cataclysmic (and/or interacting) binaries should be
present. These observational facts may indicate that NGC~6266 is now
in a dynamical state where the rate ${\cal R}_{\rm form}$ of formation
(and of hardening) of binary systems containing a neutron star and/or
a white dwarf is much larger than the rate of disruption ${\cal
R}_{\rm disr}$ of such systems.

In order to investigate in more detail the dynamical properties of
this cluster, we have used HST-WFPC2 high resolution images of the
central core and ESO-WFI wide field observations to sample the
more external regions.  The combination of these two data-sets allows us
to derive an accurate star density profile over the entire cluster
extension. \S~\ref{obs} describes the datasets and the reduction
procedures, while \S~\ref{center} presents the derived CMD. 
\S~\ref{profile} describes the determination of the cluster
center of gravity and the star density profile whereas \S~\ref{param}
is focused on the determination of the new cluster
parameters. \S~\ref{bss} is devoted to the study of the Blue Straggler
population in the central regions of NGC~6266 and, finally we discuss
the results in \S~\ref{disc}.

\section{Observations and data analysis}
\label{obs}
The photometric data used here consist of two sets.  {\it (i)---The
high spatial resolution set} includes a series of HST-WFPC2 images
obtained in August 2000, through the F555W ($V$), F336W ($U$), and
F255W ($UV$) filters as part of a long term project (GO-8709, PI:
F. R. Ferraro) aimed at studying the central stellar populations in a
sample of GCs.  In this dataset the planetary camera (PC, with the
highest resolution of $\sim 0\farcs{046}/{\rm pixel}$) is roughly
centered on the cluster center, while the WF cameras (at a lower
resolution of $\sim 0\farcs{1}/{\rm pixel}$) sample the surrounding
outer regions.  {\it (ii)---The wide field set---} secured at the 2.2m
ESO-MPI telescope at ESO (La Silla) in July 2000 using the WFI, which
has exceptional imaging capabilities by providing a mosaic of 8 CCD
chips (each with a field of view of $8'\times 16'$) for a global field
of view of $33'\times 34'$.  The cluster is roughly centered on chip
$\#2$ and observed through the $B$, $V$, $I$ broad band filters. 

\subsection{Photometry and Astrometry}  
\label{photometry}
The raw ground-based WFI images were corrected for bias and flat
field, by using standard IRAF\footnote{IRAF is distributed by the
National Optical Astronomy Observatory, which is operated by the
Association of Universities for Research in Astronomy, Inc., under
cooperative agreement with the National Science Foundation.} tools.
The point spread function (PSF) fitting procedure was performed
independently on each $B,~V$, and $I$ image, using DAOPHOT II (Stetson
1994).  A final catalog listing the instrumental $B,~V$, and $I$
magnitudes for all the stars in each field has been obtained by
cross-correlating the three catalogs.  The WFI catalog was finally
calibrated by using the data-set by Rosenberg et al. (2000) for $I$
magnitudes and by Brocato et al. (1996) for $V$ and $B$ magnitudes.
 
The photometric reduction of the high resolution HST images was carried
out using ROMAFOT (Buonanno et al. 1983\nocite{b+83}), a package
developed to perform accurate photometry in crowded fields and
specifically optimized to handle under-sampled PSFs (Buonanno \&
Iannicola 1989\nocite{b89}), as in the case of the HST-WF cameras.
PSF-fitting instrumental magnitudes have been obtained using the
standard procedure described in Ferraro et al. (1997, 2001).  The
final catalog of the F336W, F555W and F255W magnitudes was calibrated
by using the zero-points listed by Holtzman et al. (1995).  The HST
F555W band photometry has been then transformed into the Johnson $V$
system by using the stars in common between the WFPC2 and WFI
catalogs.  The photometric error of the final catalogs is dominated by
the zero-point calibration uncertainties of
$\approx\pm0.05$~mag.
 
The {\it Guide Star Catalog} ($GSCII$) was used to search for
astrometric standards in the entire WFI image field of view. Several
hundred astrometric $GSCII$ reference stars were found in each chip,
allowing an accurate absolute positioning of the sources.  An
astrometric solution has been obtained for each of the 8 WFI chips
independently, by using suitable catalog matching and
cross-correlation tools developed at the Bologna Observatory .  At the
end of the entire procedure, the rms residuals are $\approx
0\farcs2$ both in RA and Dec.

The small field of view ($2\farcm5$ on the side) of the high resolution
HST-WFPC2 images is entirely contained within the
WFI chip $\#2$, which imaged the cluster core.  We used more
than 2500 bright stars in the WFI catalog lying in the WFPC2 field of
view as {\it secondary astrometric standards}, in order to properly
find an astrometric solution for this catalog as well. We estimate
that the global uncertainty in the astrometric solution is $\le
0\farcs3$ both in RA and Dec.

A master, homogeneous catalog of magnitudes and absolute coordinates
including all the stars in the HST and the WFI catalog was finally
produced.
  
\section{The CMD of NGC~6266}
\label{center} 
\begin{figure}
\plotone{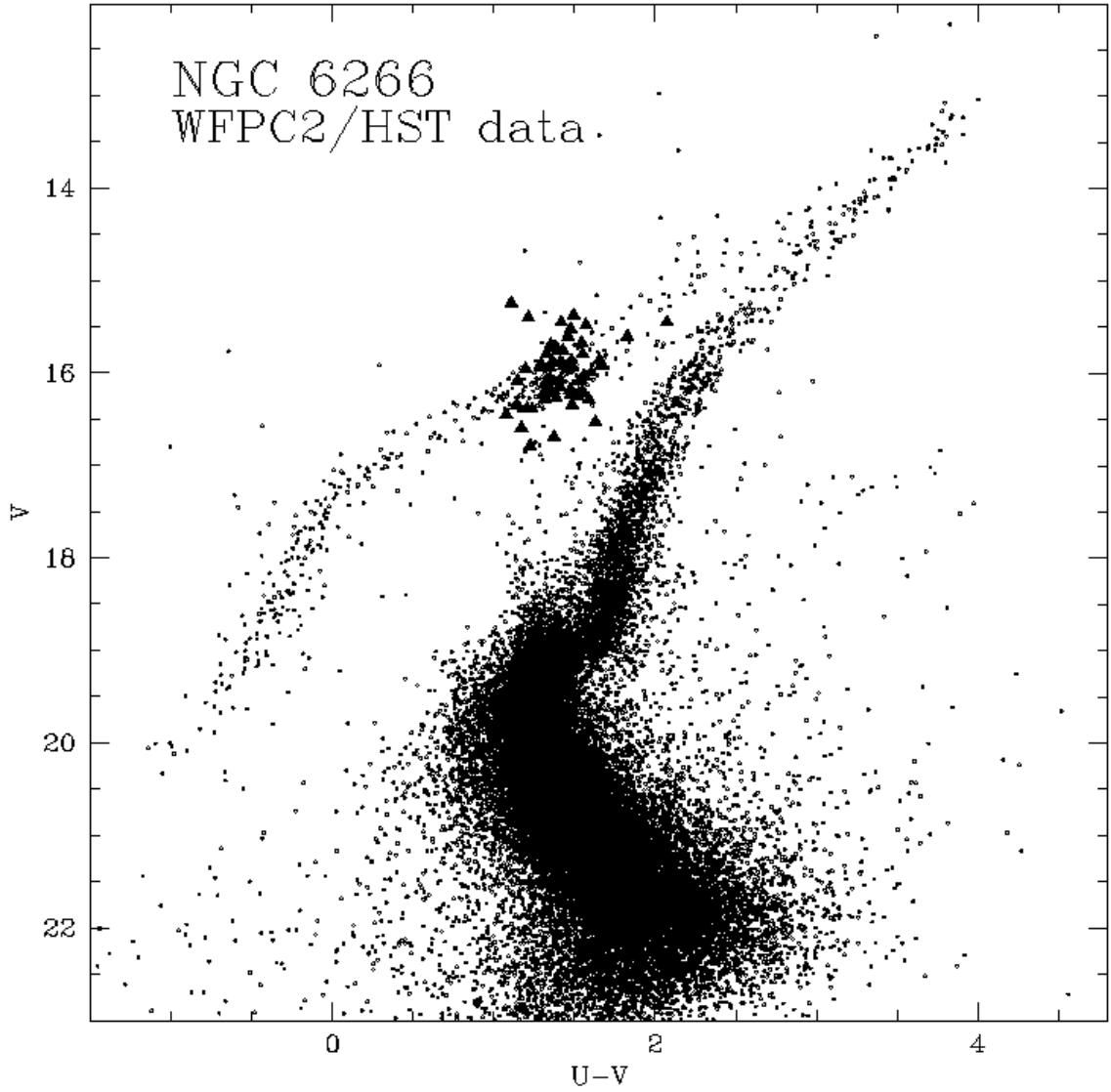} 
\caption{\label{hst} \footnotesize{CMDs
for stars in the HST/WFPC2 FoV. RRLyrae stars lying in HST/WFPC2 field
of view are marked with large filled triangles.}}
\end{figure}

Figure~\ref{hst} show the CMD in the ($V, ~U-V$) plane for the entire
HST sample. As can be seen all the cluster sequences are clearly
defined and well populated.  Particularly notable is the Horizontal
Branch (HB) morphology.  The CMD clearly shows an extended HB tail:
this feature was already suspected on the basis of previous photometry
(see for example Caloi et al. 1987, Brocato et al. 1996, and Contreras et
al. 2005) but its
true extension is revealed for the first time by the diagram shown
in Figure~\ref{hst} (see also Piotto et al., 2002).
The HB tail extends $\sim 1$ mag
below the MSTO and it is not uniformly
populated. Similar morphology has been already observed in other
clusters (NGC~2808---Sosin et al 1997; M13---Ferraro et al .  1997;
M80---Ferraro et al. 1998; NGC~6752---Ferraro et al. 2003b, see also
Catelan et al. 1998, Piotto et al. 1999 and 
Ferraro et al 1998 for a discussion on the reality of gaps along the
HB).  We defer the detailed discussion of the stellar distribution
along the HB to a future paper, where the HB
morphology of these clusters will be compared to each other.

NGC~6266 is in the sample of 62 GCs for which Ferraro et al. (1999b,
hereafter F99) derived distance modulus and photometric properties. In
doing this, F99 used the photometry presented by Brocato et al
(1996). As can be seen from their Figure 12b the HB appears quite
dispersed when compared with the CMD shown here in Figure 1. Hence,
since the zero point of the photometry presented here is based on the
photometry by Brocato et al (1996), we used our new photometry in
order to derive a more accurate level of the ZAHB
($V_{ZAHB}$). Following the procedure described in F99, we obtained
$V_{ZAHB}=16.25\pm0.10$, significantly ($\delta V=0.15$) brighter than
the values in F99. By adopting this new value, the
reddening from Harris (1996) ($E(B-V)=0.47$) and the metallicity
from Carretta\& Gratton (1997, hereafter CG97)
($[{\rm Fe/H}]_{{\rm CG97}}=-1.07$), 
we derived (from equation 4 in F99) 
$M_V^{ZAHB}=0.68$ , an apparent distance modulus of 
$(m-M)_V=15.57\pm0.15$ and a
true distance modulus of $(m-M)_0=14.11\pm0.15$ which corresponds to a
distance of 6.6 Kpc.

Recently Contreras et al. 2005 have found a large ($>200$) 
population of RR~Lyrae in NGC~6266. In order to determine the number of 
RR Lyrae variables lying in the
field sampled by the HST-WFPC2, we cross-correlated our catalog with
the list of variable stars by Contreras et al (2005). 
We identified 52 RR Lyrae in the HST/WFPC2 field of view. 
The positions of these
stars in the CMD of Figure~\ref{hst} are marked by large filled
triangles. We also identified 68 RR Lyrae in the WFI catalogue.  The
mean magnitude of all 126 identified variables is
$<V_{{\rm RRLy}}>=16.16\pm0.4$. By adopting the relation:
$$ M_V({\rm HB})=0.22({\rm [Fe/H]}+1.5)+0.56$$
 by Gratton et al
(2003) and the metallicity of NGC~6266 ($[{\rm Fe/H}]_{{\rm CG97}}=-1.07$)
 we found $M_V^{{\rm HB}}=0.65$ and an
apparent distance modulus $(m-M)_V=15.55\pm0.15$, which is fully
consistent with the value reported above.  Hence in the following we
adopt $(m-M)_V=15.57\pm0.15$, $E(B-V)=0.47$ and
$(m-M)_0=14.11\pm0.15$.
  
\section{The surface brightness and star density profile}
\label{profile} 

\subsection{The center of gravity of NGC~6266}
For the first time star counts in the central region of the cluster
can be computed using the CMD presented in Figure~\ref{hst}. Ferraro
et al. (2003b, 2004) have shown that the knowledge of the position of
individual stars in the innermost region of the cluster allows a high
precision determination of the position of the center of gravity. To
do this we applied the procedure described in Montegriffo et
al. (1995) (see also Calzetti et al. 1993) and we computed $C_{\rm
grav}$ by simply averaging the $\alpha$ and $\delta$ coordinates of
stars lying in the PC camera of the HST catalog.  In order to evaluate
any possible spurious effect due to incompleteness of the sample in the
very inner region of the cluster, we considered samples with different
limiting magnitudes ($m_{{\rm F555W}}< 20.5$, 20
respectively). For each sample we computed the barycenter of the stars
by using an iterative procedure (Ferraro et al 2003b). Both the
determinations agreed within  less than $\sim1''$.  The  
 center of gravity of the cluster 
($C_{\rm grav}$) turns out to be located at $\alpha_{\rm J2000} =
17^{\rm h}\, 01^{\rm m}\, 12\fs78,~\delta_{J2000} = -30\arcdeg\,
06\arcmin\, 46\farcs0$ with a typical $1\sigma$ uncertainty of
$0\farcs 5$ in both $\alpha_{\rm J2000}$ and $\delta_{{\rm J2000}}$,
corresponding to about $10$ pixels in the PC image.

\begin{figure}
\plotone{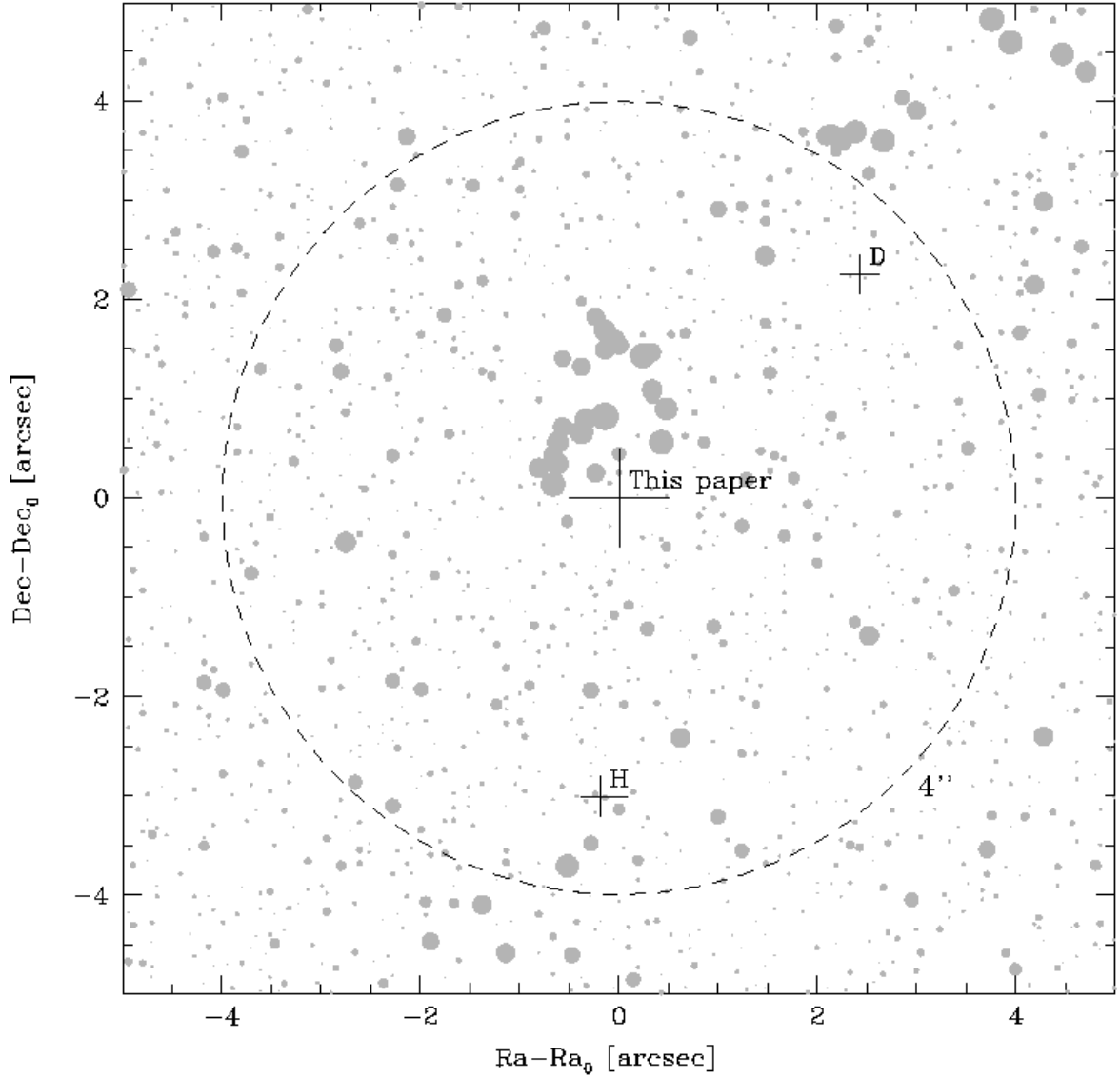} 
\caption{\label{map}
\footnotesize{Computer map of the inner 
$10\arcsec\times10\arcsec$ region of the cluster. The {\it large cross} at
$(0,0)$ coordinates indicates the adopted $C_{\rm grav}$. The
$C_{\rm lum}$ by Djorgovski (1993) and 
Harris (1996) are labeled 
with D and H, respectively.}}
\end{figure}

Figure~\ref{map} shows the $10\arcsec\times 10\arcsec$ computer map
around the cluster $C_{\rm grav}$ as determined in this work (marked
with a large cross at $(0,0)$).  The cluster luminosity center derived
by DM93 ({\it small cross} in Figure~\ref{map}) is at about $3\arcsec$
NE from our determination, whereas the center of Harris (1996)is 
about $3\arcsec$ south.

\subsection{The star density and surface brightness profile}
\label{profiles1} 

The extended data set collected for this cluster offered the
possibility to compute the radial star density profile.  As first
step, we define the star sample, paying particular care to avoid
spurious effects due to possible incompleteness. Since such effects
could be particularly important in the WFI sample, we restricted it to
stars with $r>140\arcsec$.  In the HST sample we have excluded the
external region of the sample since the particular geometry of the
WFPC2 field of view prevents an appropriate sampling of concentric
annuli used to compute the radial density profile.  A limiting
magnitude of $V\sim20.5$ was adopted for both samples. To summarize:
\begin{enumerate}
\item the WFPC2/HST sample: all stars detected in the WFPC2 catalog
with $V<20.5$ and $r<93\arcsec$ from the cluster center;
\item the WFI sample: all stars detected in the WFI catalog with
$V<20.5$ and $r>140\arcsec$ from the cluster center.
\end{enumerate}

\begin{figure}
\plotone{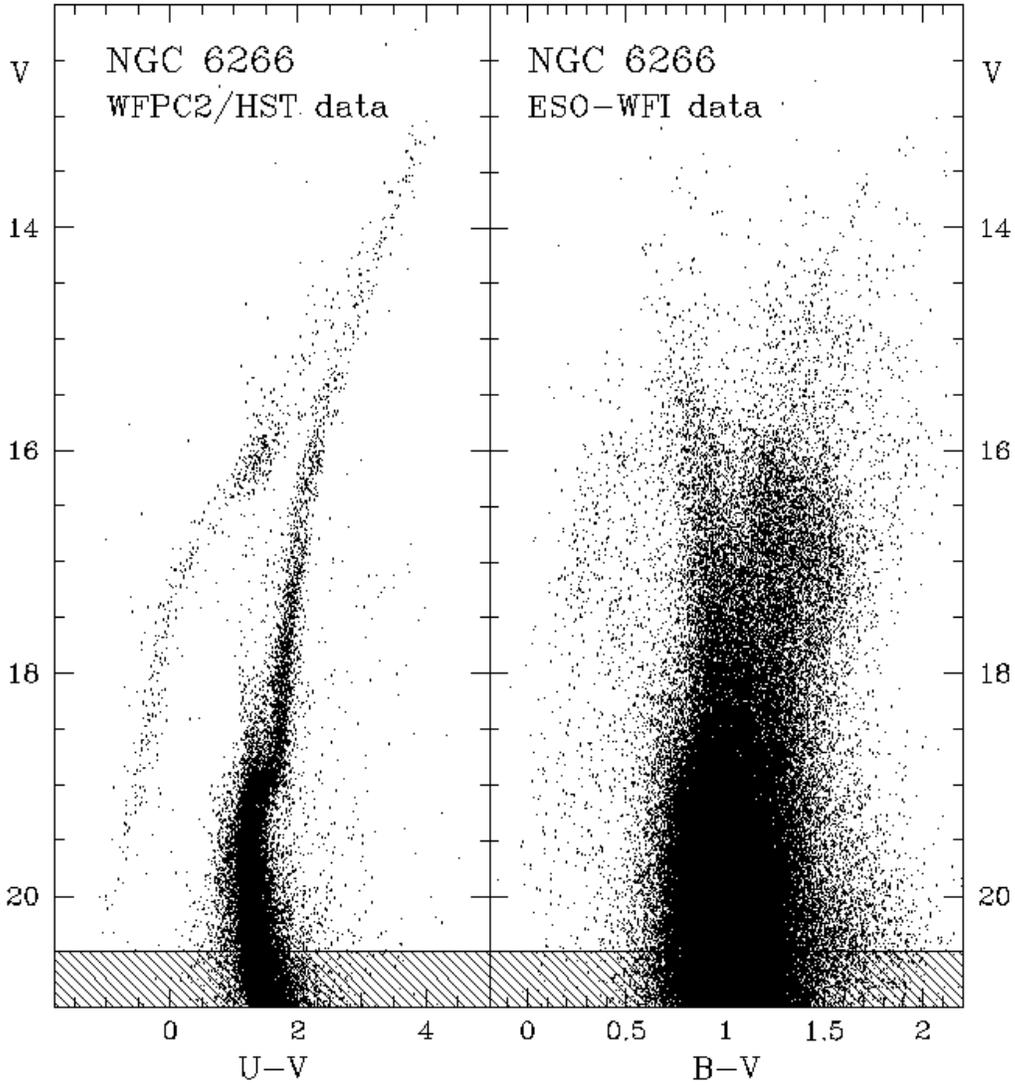} 
\caption{\label{CMD} \footnotesize{CMDs
for stars in the two samples.  {\it Left panel:} the high resolution
HST catalog ($r<93\arcsec$)in the $({\rm F555W,~F336W-F555W})$ plane. 
{\it Right panel:} 
the wide-field WFI catalog in the $(V,~B-V)$ plane,  only stars
with $r>140\arcsec$ from the cluster center are plotted. Stars with $V>20.5$
(shaded region) have been excluded from the construction of the density profile.}}
\end{figure}

Figure~\ref{CMD} shows the CMD in the ($V,~U-V$) and ($V,~B-V$) planes for
the HST and WFI samples, respectively. As can be seen, the CMD of the
WFI catalog (right panel in Figure~\ref{CMD}) is severely contaminated
by the disk population, which defines the almost vertical
blue sequence (at $0.5<B-V<1.2$) and, the bulge population at
$(B-V)\simeq1.5$, which clearly indicates the presence of metal-rich
stars. Figure~\ref{CMD} shows also a
bright portion of the cluster Red Giant Branch (RGB) at $V\sim14$ 
and $B-V\sim1.5$ and, the blue extension of the HB 
at $V\sim16$ and $B-V<0.5$.

By using the combined data set shown in Figure~\ref{CMD} we computed
the star density and surface brightness profiles, applying the standard
procedure already described in previous papers (see Ferraro et
al. 1999a, 2004). The entire photometric sample has been divided in 41
concentric annuli centered on $C_{\rm grav}$, spanning a spatial range
from $0\arcsec$ to $25\arcmin$. Each annulus has been split in a
number of sub-sectors (generally octants or quadrants, depending on
the geometry of the field covered by the HST and WFI fields). The number of
stars lying within each sub-sector was counted and star
density was obtained by dividing the average star
number by the area of the sub-sector. The
surface brightness of each sub-sector has been computed by summing the
luminosity of all the stars lying inside it, and normalizing the total
luminosity to the subsector area.

The average of the star density and surface brightness calculated in each
subsectors, at a given radius, yields the stellar
density and surface brightness of the parent annulus, respectively. 
The uncertainty in the
average values for each annulus was estimated from the variance among
the subsectors. Errors in the number counts can also be estimated from
Poison counting statistics and are consistent with the empirically
determined values.

The star density values obtained for each annulus at different
distance from the cluster center and the estimated errors are listed
in Table 1 and shown in Figure~\ref{dens1}. This is the most
complete and extended density profile ever published for this cluster,
since it samples the cluster population from the very inner core
region out to $r\sim 25\arcmin$ from the cluster center.

The computed surface brightness can suffer relatively
large fluctuations due to the small number statistics of the
bright giants. Hence, we computed three radial profiles, removing the
stars brighter than $V=12,~13,~14$, respectively.
Figure~\ref{brightness} shows that the overall structure of the
profile does not change with the adopted magnitude limit.  In
contrast, the dispersion among different sub-sectors significantly
decreases once the brightest stars are excluded.

However, the surface profile shown in Figure~\ref{brightness}
indicates a central brightness of $\mu_{v,0}\sim 15.2$
mag\,arcsec$^{-2}$. This value is fully consistent with that listed by
Harris (1996) ($\mu_{v,0}\sim15.35$ mag\,arcsec$^{-2}$) and DM93
($\mu_{v,0}\sim15.19$ mag/arcsec$^2$), hence in the following we adopt
this value for the central surface brightness of the cluster.

\subsection{The dynamical state of NGC~6266}
\label{profiles2} 

The shape of the radial density profile of GCs is generally used to
infer the dynamical state of the system (see the pioneering work by King
1966 and Djorgovski \& King 1986). In particular, Djorgovski \& King
(1986) defined two classes of clusters: the so-called King Model (KM)
cluster and the PCC cluster, depending of the
central density profile. KM clusters have density profiles which can be
reproduced by a single-component King Model (King 1966) with a flat
isothermal core and a steep envelope characterized by two parameters:
the core radius ($r_c$) and the tidal radius ($r_t$) or,
alternatively, the concentration $c=\log(r_t/r_c)$. The star
density profile of a PCC cluster follows an almost pure power law with
an exponent $\sim1$.  However Meylan \& Heggie (1997) stated that the
general dynamical status of a cluster can be deduced directly from the
concentration value ``c'', since all clusters with high concentration
parameter value ($c>2$) can be considered as collapsed or on the verge
of collapsing.

\begin{figure}
\plotone{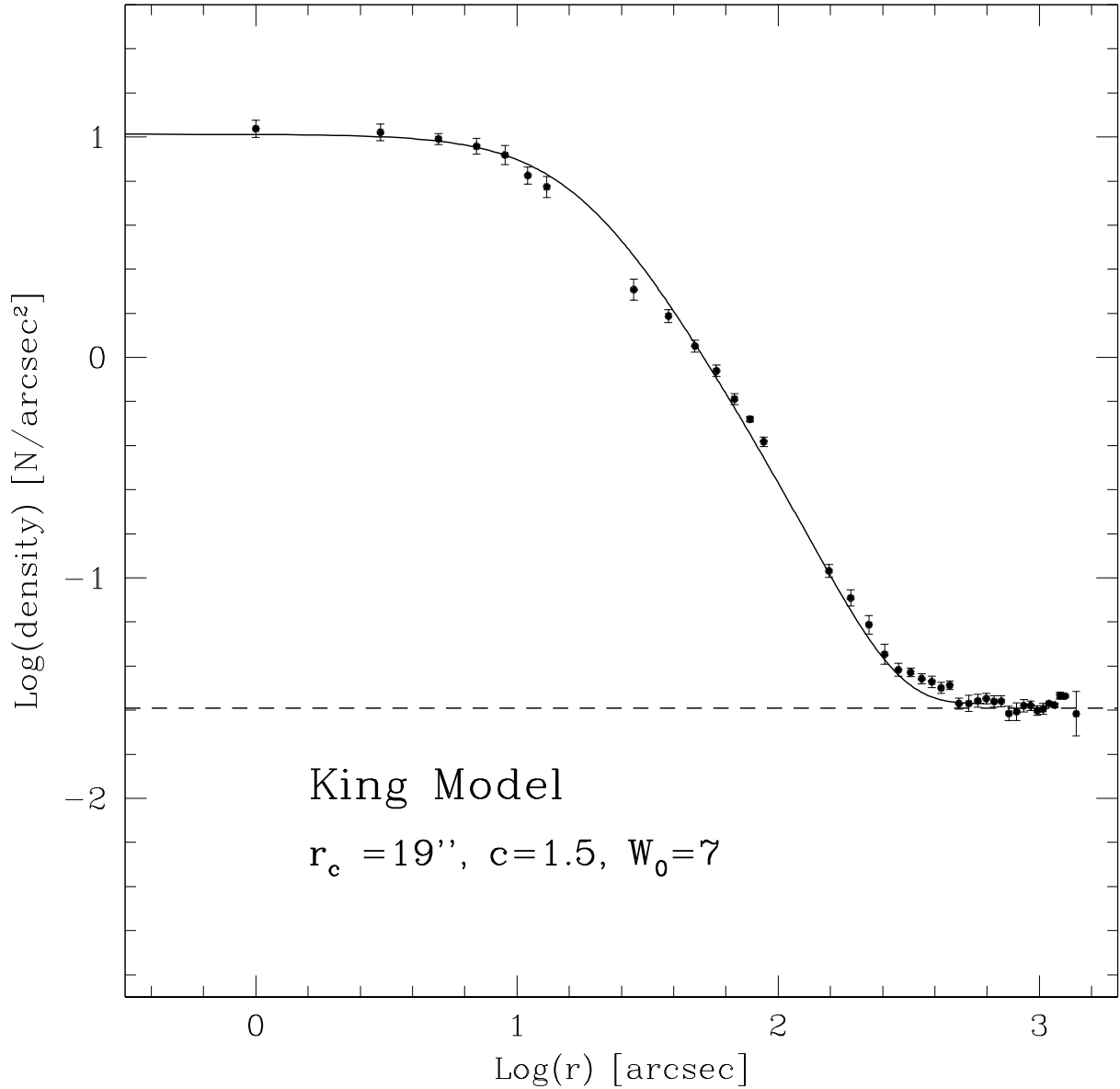} 
\caption{\label{dens1} \footnotesize{Observed radial density profile
with respect to the the adopted $C_{\rm grav}$.  The solid line is the
best fit King model ($c=1.5$) to the observed density profile over the
entire extension. A value of $\sim92~{\rm stars\,arcmin}^{-2}$
is representative field contamination (dashed horizontal
line). The adopted value for the parameter $W_0$ (the central
potential parameter as defined by King 1996) is also reported.}}
\end{figure}

\begin{figure}
\plotone{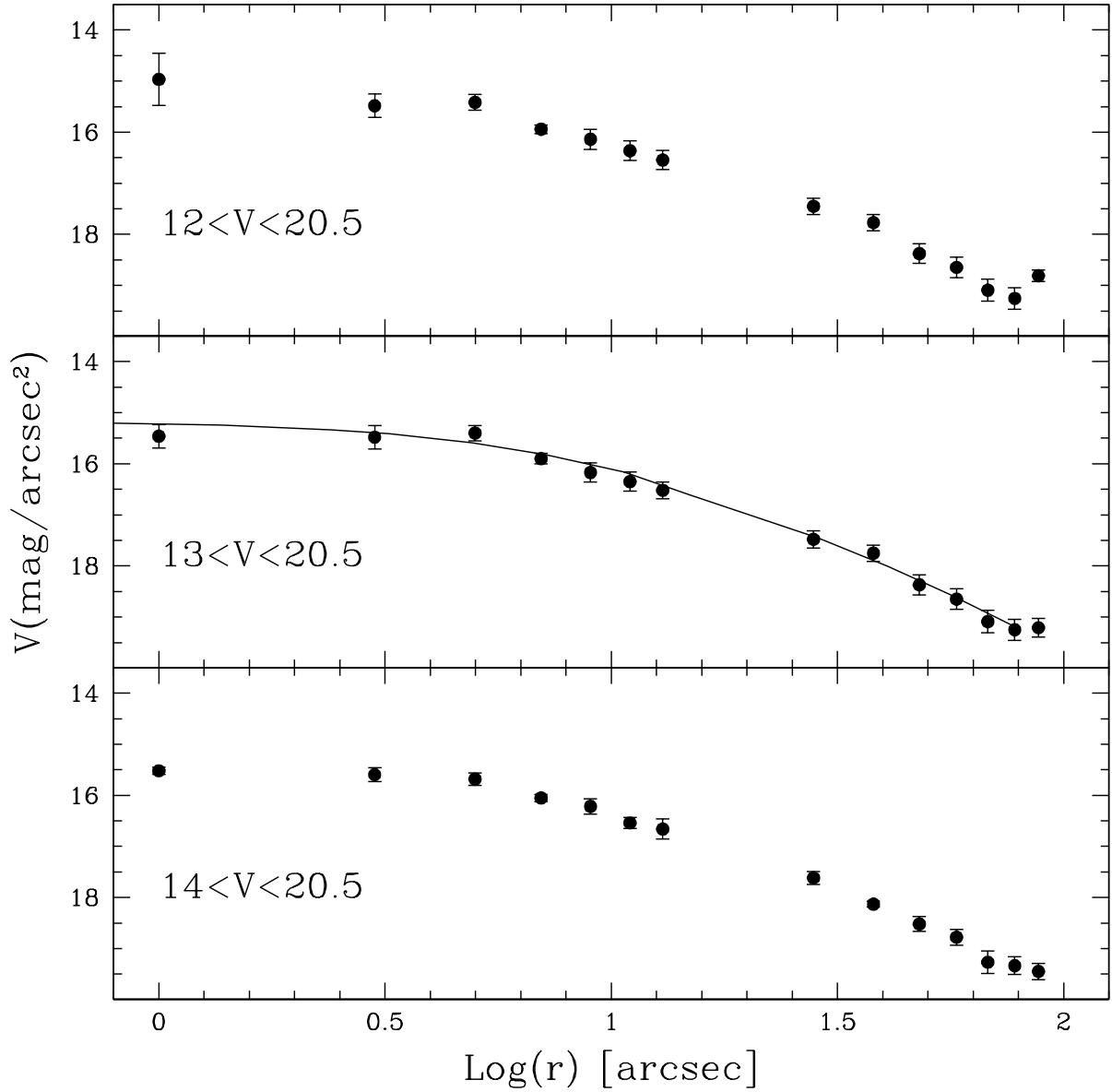} 
\caption{\label{brightness}
\footnotesize{Radial brightness profiles computed, for 
the WFPC2/HST sample, by removing the stars
brighter than $V=12,13,14$, respectively, the faintest limit is
$V=20.5$ (see Figure~\ref{CMD}). The solid line is the best-fit solution.}} 
\end{figure}

In order to model the observed profile, isotropic, single-mass
King-models projected onto the cluster area have been computed by
using the Sigurdsson \& Phinney (1995) code. As is shown in
Figure~\ref{dens1}, the observed radial profile is well reproduced by
a KM with an extended core ($r_c=19\arcsec$). Note that the value of
the core radius is significantly larger (almost by a factor of 2) and
the concentration ($c=1.5$) significantly lower than $r_c=10\farcs 7$
and $c=1.7$ tabulated by Trager et al. (1995).  Since the star density
profile is not affected by statistical fluctuations due to the
presence of a few bright giants, it offers the best route for deriving
cluster parameters (see also Lugger, Cohn \& Grindlay 1985 and
Ferraro et al 2003b).

Both the existence of an extended core and a relatively low value of
the concentration parameter suggest that the cluster has not yet
experienced the collapse of the core. Hence, the first result of this
study is that {\it NGC~6266 is not a PCC cluster}.

The star density profile shown in Figure~\ref{dens1} also allows us to
evaluate the contribution of the background stellar population, which
is substantial due to the fact that the cluster is located in the
direction of the Galactic Bulge. Our star counts can be used to
evaluate the projected Bulge star density down to $V\sim20.5$.  By
averaging star counts in the region where the radial profile become
flat (corresponding to $r>470\arcsec$ from the cluster center), we
have estimated a value of $\sim92$ ${\rm stars\,arcmin^{-2}}$. The
background level is shown as a dashed horizontal line in
Figure~\ref{dens1}. The poorer fit of the King Model to the cluster
in the region of transition to the background could arise from our
modeling of the background as spatially uniform. 

\section{The new cluster parameters}
\label{param}
Since the KM fitting of the observed stellar density profile of the
cluster suggested values of $r_c$ and $c$ significantly different from
those previously measured, we used them to re-determine the cluster
structural parameters. In doing this, we adopted the equations given by
Djorgovski 1993 (from here D93). Following his assumptions the central
luminosity density $\rho_0^L$ is:

$$\rho_0^L=\Sigma_0/(r_c~p)$$ 

\noindent where $r_c$ is in parsecs and
$\Sigma_0$ is the central surface brightness in $L_{V\odot}\,{\rm pc}^{-2}$,
evaluated as follows:

$$\log(\Sigma_0)=0.4\, [26.362-\mu_{V,0}(0)]$$ 

\noindent and $p$ is a function
which depends on the cluster concentration ($c$), given by the following
expression:

$$\log(p)=-0.603\times10^{-c}+0.302$$ 

\noindent and $\mu_{V,0}(0)$ is the central
brightness of the cluster corrected for extinction.

In computing these quantities we adopted the distance modulus and the
reddening discussed in Section 3.  The resulting central luminosity
density is {$\log(\rho_0^L)=4.98~[L_{\odot}\,{\rm pc}^{-3}]$}.  If,
according to D93, we use a mass-to-light ratio $(M/L_V)=3$ for
converting the central luminosity density into the central mass
density, we derive a central density value of $\log
\rho_0=5.46~[(M_{\odot}\,{\rm pc}^{-3})]$, which has to be compared with
the value listed in D93 $\log \rho_0=5.7~[(M_{\odot}\,{\rm pc}^{-3})]$.
All the derived parameters are listed in Table~\ref{6266dj} along with
the values by DM93 for comparison.  

To homogeneously compare the properties of NGC~6266 with those of other
GCs, we also calculated new values of $\rho_0$ for a sample of GCs
previously studied with similar techniques (see F03, Sabbi et al. 2004,
Ferraro et al. 2004).  In doing this we adopted the distance scale
defined by F99 (also the adopted reddening is from Table 2 by F99) and
the $r_c$ derived from the radial density profile of each cluster (see
F03, Mapelli et al. 2004, Ferraro et al. 2003b). Results are listed in
Table~\ref{clusts} along with  the previous
determinations by D93. The comparison shows that the different distance
modulus adopted here (with respect to D93) has a direct effect on the
derived value of the cluster central density, with a major impact on
the derived binary-binary collision rate (see Section~\ref{disc}).

\section{The population of Blue Straggler Stars in the core of NGC~6266}
\label{bss}

Special classes of stars, in particular those resulting from
the evolution of binary systems, can be used as tracers of the GC
dynamical evolution. This is the case of the BSS. 
From an observational point of view, the high angular
resolution and UV imaging capabilities of HST has allowed
searches for BSS candidates in the core of highly concentrated GCs
leading to a burst of activity in this sector (see for
example Ferraro et al. 1997, 1999a, F03; Piotto et al. 2004).

\begin{figure}
\plotone{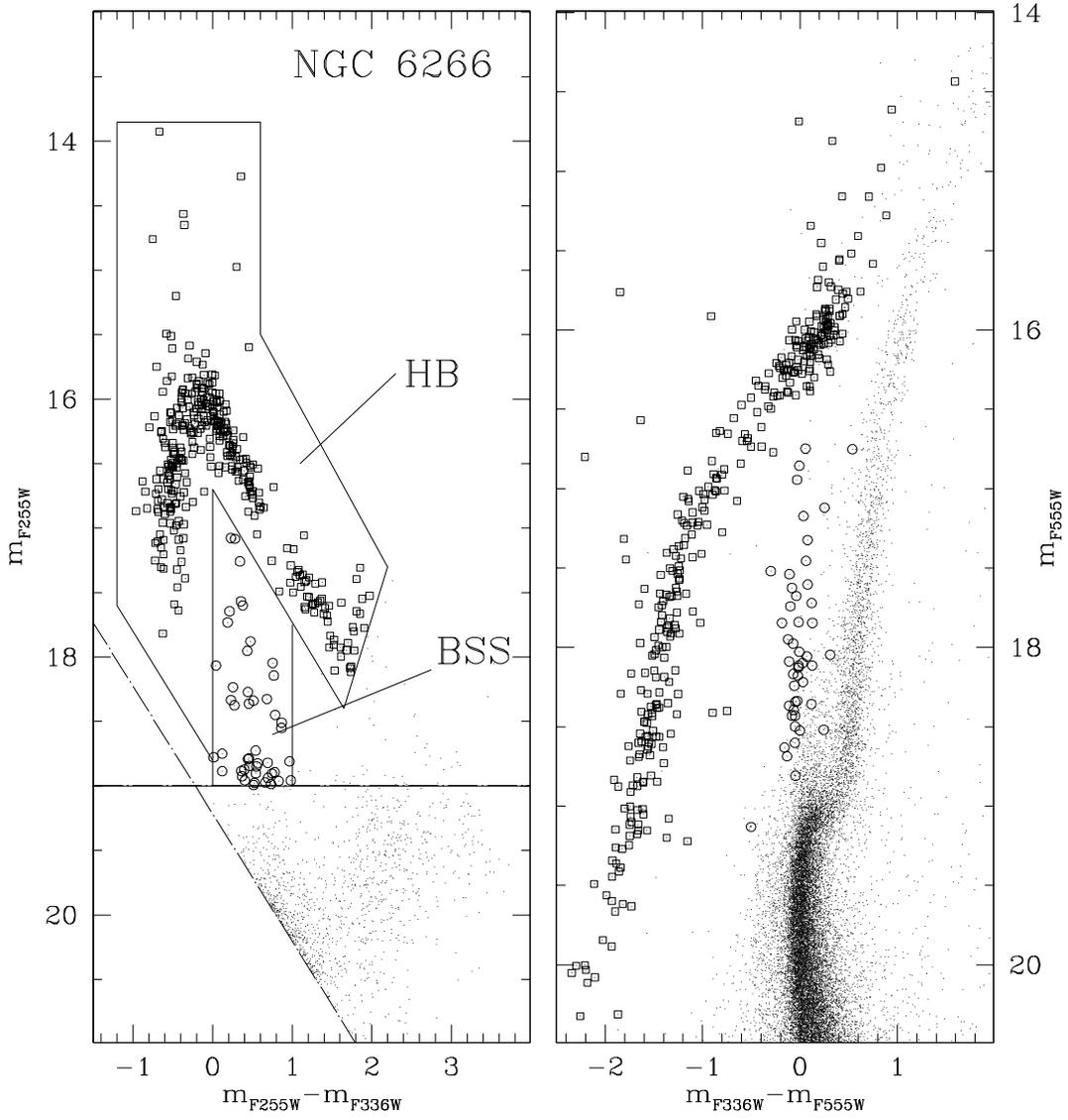} 
\caption{\label{new}
\footnotesize{BSS (open circles) and HB (open squares) in the 
$({\rm F255W,~F255W-F336W})$ CMD of
NGC~6266 ({\it left panel}) and $({\rm F555W,~F336W-F555W})$ 
CMD ({\it right panel}). Note that only non variable HB
stars are plotted.
The adopted selection boxes 
in the UV-CMD are shown.}}
\end{figure}

An obvious difficulty in studying BSS is the definition of an unequivocal
selection criterion since the BSS sequence generally merges into the
normal cluster MS without any discontinuity. However, previous work
done by our group (Ferraro et al. 1997, 1999a, 2001, F03,
2004) has shown that the UV plane (the $m_{255}, m_{255}-m_{336}$ plane in
particular) is ideal for selecting BSS. In this plane the BSS
sequence appears almost vertical and significantly brighter than the
TO and Sub Giant Branch (SGB).  In order to properly compare the
BSS population in different clusters, here we follow the same
procedure defined in Ferraro et al. 1997 and adopted in F03. A
limiting magnitude $m_{F255W}=19$ has been adopted in selecting
bright BSS once the CMD of each cluster has been shifted to match that
of M3.  Following this criterion, 47 BSS candidates have been selected
in NGC~6266.  Figure~\ref{new} ({\it Left panel}) shows the selection box
adopted for the BSS in the UV plane. The position of the
selected BSS in the $(V,U-V)$-CMD is also shown in the {\it
Right panel} of Figure~\ref{new}.

Since BSS are the result of two stars merging in a binary system, one can
expect that in a dynamically relaxed stellar system, BSS are
centrally concentrated with respect to {\it normal} stars (see Bailyn
1995) because of mass segregation. Indeed, the central relaxation time
of even a sparse cluster is significantly lower than the lifetime of a
1--1.5 $M_{\odot}$ MS star. Mass segregation has been
observed in all the GCs properly observed up to now (see F03 and
reference therein) except $\omega$~Centauri 
(Ferraro et al. 2005), which may not be a
true GC (Tsuchiya et al., 2004).

When using UV-CMDs the obvious reference population is the
HB, which appears quite luminous and well defined. HB stars have masses
($\sim0.5-0.8~M_{\odot}$) significantly smaller than those of BSS and
their lifetimes are typically shorter than GC relaxation times.
Figure~\ref{new} ({\it Left panel}) shows the adopted selection
box for HB stars in the UV plane. Indeed they are
easely separable from the cooler (hence fainter) 
RGB stars. The {\it right panel} of Figure~\ref{new} shows
their position in the (V,U-V) CMD. As can be seen, the
selection boxes defined in the UV plane assure the proper
inclusion of  the entire HB populations down to the
faintest end of the branch.
In the WFPC2 field of view, we have identified a
total  population of 395 HB stars (52 RRLyrae - see Section 3 and
343 non variable stars, respectively)\footnote{Note that
only non-variable HB
stars have been plotted in both panels of Figure~\ref{new}}. 

\begin{figure}
\plotone{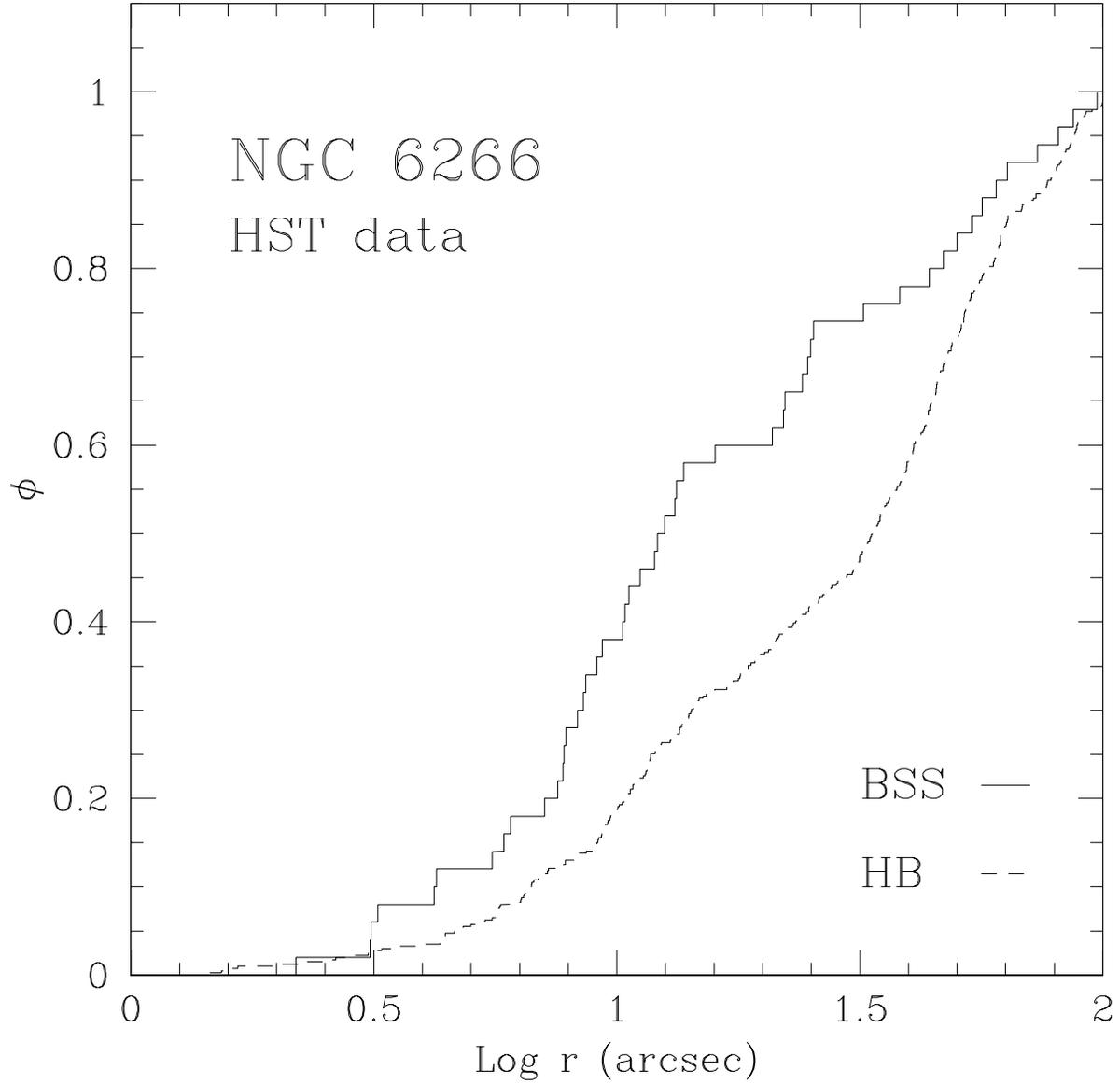} 
\caption{\label{distr}
\footnotesize{Cumulative radial distribution of BSS (solid line) with respect
to the HB stars (dashed line) as a function of their projected distance ($r$)
from the cluster center. The probability that the two populations are extracted
from the same distribution is $P=2\times10^{-4}$}}
\end{figure}

We have determined the cumulative radial
distribution of BSS and compared to HB stars as a function of the
projected distance from the cluster center. Figure~\ref{distr} shows
the result. It is evident from the plot that BSS (solid line) are
significantly more concentrated toward the center than HB stars. We
applied a Kolmogorov-Smirnov test to both distributions to check
the statistical significance of the detected difference: the
probability that the BSS population and the HB stars are extracted
from the same {\it parent} distribution is $P=2\times10^{-4}$. Thus,
the central concentration of the BSS is confirmed with a $>99\%$
statistical significance.

\subsection{BSS Specific Frequency}

\begin{figure}
\plotone{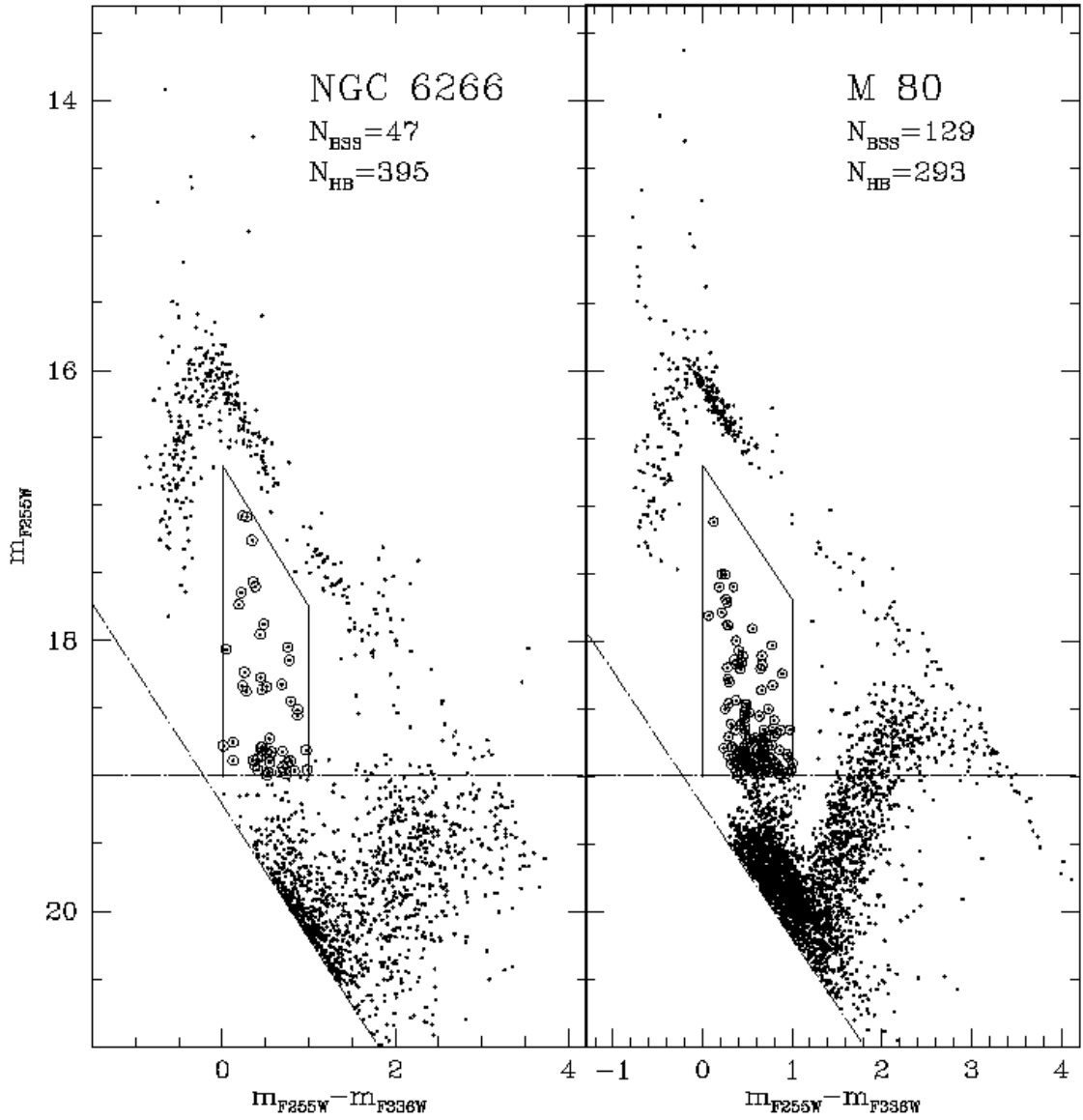} 
\caption{\label{bss1}
\footnotesize{BSS (open circles) in the $({\rm F255W,~F255W-F336W})$ CMDs of
NGC~6266 ({\it left panel}) and M80 ({\it right panel}) for comparison. 
The CMDs have been shifted as suggested in F03. The horizontal line 
shows the limiting magnitude of the adopted selection.}}
\end{figure}

In order to make a direct comparison with the BSS population in the
previous published GCs (Ferraro 1997, F03, Sabbi et al 2004, etc), we
computed the BSS specific frequency. As an example,
Figure~\ref{bss1} shows the UV-CMD of NGC~6266 (left
panel) compared to that of M80 (right panel) presented in F03.
From Ferraro et al. (1999a), the
specific frequency ($F^{\rm BSS}_{{\rm HB}}$) is defined as the ratio
between the number of BSS ($N_{{\rm BSS}}$) to the number of HB 
($N_{{\rm HB}}$) stars. It's worth noticing that the number of stars in each
post-MS phase is an excellent indicator of the sampled cluster
luminosity (see Renzini \& Buzzoni 1986).  Hence $F^{\rm BSS}_{{\rm
HB}}$ is a suitable tool for a quantitative comparison between BSS
populations in different GCs. A further advantage is that $F^{\rm
BSS}_{{\rm HB}}$ is a purely observational quantity and can be easily
computed in the UV-CMDs, where the HB population is well separated
from the other sequences. The values of $F^{\rm BSS}_{{\rm HB}}$
obtained  in 6 GCs by F03, in NGC~6752 by Sabbi et
al. 2004, in 47 Tucanae by Ferraro et al. 2004 and the
one calculated in NGC~6266 are reported in Table~\ref{bss_freq},
sorted according to increasing $F^{\rm BSS}_{{\rm HB}}$ (the values of
$N_{{\rm BSS}}$ and $N_{{\rm HB}}$ are also reported).  All the
clusters listed in Table~\ref{bss_freq} have been observed 
in the central regions by using the WFPC2/HST and 
hence represent an homogeneous sample for the
sake of comparison. 

We found a particularly low BSS specific frequency
($F^{\rm BSS}_{{\rm HB}}=0.13$) in NGC~6266, a value that lies between that of
M13 and 47Tuc. It is interesting to notice that NGC~6266 is the highest
central density cluster in the sample but it has one of the lowest BSS specific
frequencies. On the other hand F03 noticed that NGC~288, the cluster
with the lowest central density in their sample, shows the largest BSS
specific frequency.  This is further evidence that the central
density of a cluster is not the driving factor in determining the number
of BSS (see also  Davies et al., 2004).

\section{Discussion and Conclusions}
\label{disc}
 
By adopting the cluster parameters computed in Section~\ref{param}, we
can now investigate some dynamical properties of NGC~6266.  
We computed the mean time interval between single-single ($t_{ss}$) and
binary-binary
($t_{bb}$) collisions using equations 13 
and 14 respectively from Leonard (1989):
 
$$
t_{ss}=5.5\times10^9\left(\frac{1\,{\rm pc}}{r_c}\right)^{3}
\left(\frac{10^3\,{\rm pc}^{-3}}{n_0}\right)^2\times
\left(\frac{V_{rms}}{5\,{\rm km\,s}^{-1}}\right)\left(\frac{0.5M_{\odot}}{m_{*}}\right)
\left(\frac{0.5R_{\odot}}{R_{*}}\right)[yr],
$$

$$
t_{bb}=1.7\times10^7\left(\frac{1\,{\rm pc}}{r_c}\right)^{3}
\left(\frac{10^3\,{\rm pc}^{-3}}{n_0}\right)^2\times
\left(\frac{V_{rms}}{5\,{\rm km\,s}^{-1}}\right)\left(\frac{0.5M_{\odot}}{m_{*}}\right)
\left(\frac{1AU}{a}\right)[yr],
$$ 

\noindent where $a$ is the initial semimajor axis of each binary which can be
estimated for each cluster according to the relation (F03)
$$
a=\frac{GM}{9\sigma^2_0},
$$
$n_0$ is the central number density (number of stars per parsec$^3$)
and $V_{rms}$ is the relative root-mean-square velocity, which can be
approximated with the central velocity dispersion ($\sigma_0$, see
Table 5). Leonard (1989) derived his equation (14) under the assumption
that the binary frequency is 100\% in the core and the star mass is
$m_{\ast}=0.5M_{\odot}$. Assuming an average mass of
$m_{\ast}=0.2M_{\odot}$ (Kroupa 2001) as suggested in F03, 
and $R_{\ast}=0.5R_{\odot}$, we
computed the number of single-single ($ss$) and binary-binary 
($bb$) encounters occurring per
Gyr ($N_{ss,bb}$) in the core of NGC~6266 to be:
$$
N_{ss,bb}=\frac{10^9}{t_{ss,bb}} \;\;\;\;\;\;   [{\rm encounters\,Gyr^{-1}}].
$$ 
In Table~\ref{bb} we give the results obtained for NGC~6266, 
compared to the 8 GCs which have been previously
searched for BSS in the central regions by using the same
observational strategy.  The values of the encounter rate
are normalized to those obtained for 47 Tuc.  
NGC~6266
has a predicted {\it{bb}} encounter rate $N_{bb}$
which is much greater than for any other analyzed GC. As shown in
Table \ref{bb}, the same result holds for the
close encounters between two single stars leading to tidal capture. 
Both of these classes of
dynamical interactions (together with the binary-single 
[$bs$]
encounters)
may lead to the merging of the two
stars, thus potentially producing a BSS.  Since all collision modes should be
effective in increasing the population of BSS, one would expect to see
a relatively larger specific frequency of BSS in NGC~6266. This
expectation is not born out by the results
of Table \ref{bss_freq}. The discrepancy might be reduced
if the binary fraction in NGC~6266 were significantly smaller than in
the other GCs, but this hypothesis can hardly be reconciled with the
estimated high rate of binary formation ${\cal R}_{\rm form}$ computed
for this cluster (see later).  Perhaps the results given in 
Tables \ref{bss_freq}
and \ref{bb} can be understood if the BSS in NGC~6266 result
from primordial binaries which have survived in the outer part of the
cluster eventually wandering into the center where interactions drive
them to merge (as we suggested for 47~Tuc, see Ferraro et al. 2004).

On another side, the very high rate of binary-binary and binary-single
star interactions in NGC~6266 may well explain why all of its
known MSPs are in binary systems.  In fact, P03
suggested that NGC~6266 is in a particular dynamical phase in which
the rate of binary system formation ${\cal R}_{\rm form}$ is larger
than their destruction rate (${\cal R}_{\rm disr}$). Following the
prescription of Verbunt (2003), we recomputed those ratios for all
clusters listed in Table~\ref{coll}.  According to Verbunt (2003)
${\cal R}_{\rm form}\propto\rho_0^{1.5}r_c^2$ and ${\cal R}_{\rm
disr}\propto\rho_0^{0.5}r_c^{-1}$, where $\rho_0$ is the central
density (expressed in $M_{\odot}\,{\rm pc}^{-3}$) and $r_c$ is the core radius
of the cluster. In Table~\ref{coll} we give the values of ${\cal
R}_{\rm form}$ and ${\cal R}_{\rm disr}$ normalized to 47 Tuc (as done
by P03). A quite useful quantity is the ratio ${\cal R}_{\rm
form}/{\cal R}_{\rm disr}$ which quantifies the binary survival rate
in each environment. NGC~6266 shows, by far, the
largest value of this ratio among the clusters listed in
Table~\ref{coll}. The 2nd ranked cluster is M3 in which all the
MSPs discovered so far are also in binaries.

In summary, we have been able to recompute the ratio
${\cal R}_{\rm form}/{\cal R}_{\rm disr}$ and to show that {\it (i)} its
high value nicely agrees with the hypothesis (supported by
observations of MSPs and X-ray sources) that NGC~6266 has experienced
(or is experiencing) a phase of very high production of binary
systems. Additionally, we show that {\it (ii)} the highly
effective binary production phase is occurring while the cluster has
not yet undergone the collapse of the core, and that {\it
(iii)} the formation channel that produces a wealth of binary systems
hosting neutron stars or white dwarfs is not efficient in producing
BSS.

We can also compare NGC~6266 with the sample of GCs searched for BSS
with an approach similar to that used in this work. In particular,
inspection of Tables \ref{bss_freq} and \ref{coll} shows that in GCs
hosting MSPs the specific frequency of BSS never exceeds 0.28. This
sample includes the post-core collapsed globular cluster, NGC~6752, in
which only a few BSS have been found (Sabbi et al. 2004). Conversely,
up to now no MSPs have been found in M80 and NGC~288, the two clusters
with the largest specific frequency of BSS measured so far (probably
generated by different channels: primordial binaries in NGC~288 and
collisional binaries in M80).  The difficulty in
modeling the observational biases involved in discovering radio
pulsars in GCs prevents any firm conclusion to date, but an
anticorrelation between MSP content and BSS specific frequency seems
emerging. GCs with large populations of BSS seem to have a
small population of MSP.

\acknowledgements{\small Financial support to this research has been
provided by the {\it Ministero dell'Istruzione, dell'Universit\`a e
della Ricerca} (MIUR) under the grant {\it PRIN-2003023549}.
We also acknowledge the financial contribution from contract
{\it ASI-INAF I/023/05/0}.
We warmly thank M. Catelan for a careful reading of this paper and for
helpful discussions. We also thank the referee for useful
comments that improved the paper presentation.}

\newpage
\newdimen\minuswidth    
\setbox0=\hbox{$-$}
\begin{deluxetable}{cccc|cccc}
\scriptsize \tablewidth{12cm}  
\tablecaption{\label{t:mjd}
Surface density in different annuli around the NGC~6266 center.}
\tablecomments{$r1$ and $r2$ are the inner and outer radii
  of each annulus in arcsec.}
\startdata \\
\hline
\hline
       &	&                            &     & &      &     
                            &     \\
  $r1$  &    $r2$  &  $ \frac{N}{{\rm arcsec^2}} $ & err &  $r1$  &  $r2$  & 
  $ \frac{N}{{\rm arcsec^2}} $ & err \\
       &	&                            &     & &      &      
                            &     \\
\hline        			 
     0.0  &   2.0   &    10.89 & 0.04&  371.0  &   404.0  &   0.034 & 0.026     \\
     2.0    &   4.0  &   10.49  &0.04&  404.0  &   437.0  &   0.032 & 0.025  \\
     4.0    &   6.0  &  9.80   & 0.02 &  437.0   &   470.0 & 0.032 &  0.020 \\
     6.0    &   8.0  &  9.07  &  0.03&   470.0  &   514.5  & 0.027 &  0.025  \\
     8.0   &   10.0   & 8.28 &   0.04&  514.5  &   559.0  &  0.027 &  0.036  \\
    10.0    &  12.0  &  6.69 &   0.04 &  559.0  &   603.5  &  0.028 &  0.029  \\
    12.0    &  14.0   & 5.93  &  0.05 &  603.5  &   648.0  & 0.028 &  0.023	\\
    23.0    &  33.0   & 2.03 &   0.05 &  648.0  &   692.5  &   0.027 & 0.026	  \\
    33.0    &  43.0   & 1.54  &  0.03 &  692.5  &   737.0  &	0.027 &0.024	  \\
    43.0    &  53.0  &  1.13 &   0.03 &  737.0  &   791.0  &	0.024 &0.032	  \\
    53.0     & 63.0  & 0.869 &   0.025 & 791.0  &   845.0  &    0.024 & 0.004	  \\
    63.0    &  73.0  & 0.647 &   0.025 & 845.0  &   899.0  &    0.026 & 0.003	 \\
    73.0    &  83.0  & 0.524 &   0.012& 899.0	&   953.0  &   0.026 & 0.021	 \\
    83.0    &  93.0  & 0.415 &   0.021& 953.0  &  1007.0  &    0.025 & 0.021	 \\
   140.0  &   173.0  &  0.107  & 0.003&     1007.0 &  1061.0  &0.025 & 0.024	 \\
   173.0  &   206.0  & 0.081  &  0.037&   1061.0 &  1116.0  &	0.027 &0.009	\\
   206.0  &   239.0  & 0.061  &  0.042&   1116.0 &  1171.0  &	0.027 &0.008	\\
   239.0  &   272.0  &  0.045 &0.046&    1171.0 &  1226.0  & 0.029 & 0.015  \\
   272.0  &   305.0  & 0.038  & 0.003&   1226.0 &  1281.0  &	0.029 &0.008	\\
   305.0  &   338.0  & 0.037  & 0.002&   1281.0 &  1486.0  &	0.024 &0.101	 \\
   338.0  &   371.0  &  0.035 & 0.022&& & & \\

\hline
\enddata
\end{deluxetable}

\newpage
\newdimen\minuswidth    
\setbox0=\hbox{$-$}
\begin{deluxetable}{cccccccc}
\scriptsize \tablewidth{14cm}  
\tablecaption{\label{6266dj}
NGC~6266 parameters}
\startdata \\
\hline
\hline
                                 \\
  &  E(B-V) &  $(m-M)_0$ & $\mu_v(0)$ & $\mu_{0,v}$ & $r_c$ & c & $ Log(\rho_0)$ \\
  & & & & & [arcsec] & & $[M_{\odot}{\rm pc}^{-3}]$ \\							
\hline 
 DM93  & 0.50   & 13.70  & 15.2 & 13.64  & 10.7 & 1.8 & 5.7 \\
 This Paper     & 0.47   &  14.11 & 15.2 & 13.74  & 19   & 1.5 & 5.46 \\
\hline
\enddata
\tablecomments{$E(B-V)$ from Harris 1996}
\end{deluxetable}

\newdimen\minuswidth    
\setbox0=\hbox{$-$}
\begin{deluxetable}{cccccccc}
\scriptsize \tablewidth{15cm}  
\tablecaption{\label{clusts}
Structural paramenters for a sample of clusters searched for BSS}
\startdata \\
\hline
\hline
                                 \\
Cluster &[Fe/H]& E(B-V) & $(M-m)_0$ & Dist & $r_c$  &\multicolumn{2}{c}{$Log(\rho_o)$} \\
  &  & & & [Kpc] & [arcsec]  & \multicolumn{2}{c}{$[M_{\odot}{\rm pc}^{-3}]$} \\	
  & & & & & &This Paper &DM93   \\
\hline 
 M 3          & -1.34 & 0.01 & 15.03 & 10.1 & 30  & 4.0 & 3.5  \\
 M 13	      & -1.39 & 0.02 & 14.43 & 7.7  & 40  & 3.9 & 3.4 \\
 M 80	      & -1.41 & 0.18 & 14.96 & 9.8  & 6.5 & 5.4 & 5.4   \\
 M 10	      & -1.41 & 0.28 & 13.38 & 4.7  & 40  & 4.0 & 3.8  \\
 NGC~288      & -1.07 & 0.03 & 14.73 & 8.8  & 85  & 2.3 & 2.1        \\
 M 92	      & -2.16 & 0.02 & 14.78 & 9.04 & 14  & 4.7 & 4.4   \\
 NGC~6266     & -1.07 & 0.47 & 14.11 & 6.6  & 19  & 5.5 & 5.7     \\
 NGC~$6752^{\ast}$   & -1.42 & 0.04 & 13.18  & 4.3  & 5.7  & 5.6 & 5.2 \\
 NGC~104      & -0.70 & 0.04 & 13.32 & 4.6  & 21  & 5.3  & 5.1 \\
 
\hline
\enddata
\tablecomments{$\ast$ Post core collapse cluster.}
\end{deluxetable}

\newdimen\minuswidth    
\setbox0=\hbox{$-$}
\begin{deluxetable}{ccccc}
\scriptsize \tablewidth{8cm}  
\tablecaption{\label{bss_freq}
BSS population in a sample of GGCs.}
\startdata \\
\hline    
\hline
                                 \\
  Cluster  &    $N_{{\rm BSS}}$ & $N_{{\rm HB}}$ &  $F^{\rm BSS}_{{\rm HB}}$ &  $r_{1/2}^{\rm BSS}/r_c$  \\
  							\\
\hline 
 M13           & 16   & 228  & 0.07 & 1.15  \\
 NGC~6266      & 47   & 395  & 0.12 & 0.63  \\  		 
 NGC~104$\dag$ & 53   & 314  & 0.17 & 0.87  \\       
 NGC~$6752^{\ast}$ & 28   & 156  & 0.18 & 0.82  \\
 M10           & 22   & 81   & 0.27 & 0.85  \\
 M3            & 72   & 257  & 0.28 & 0.73  \\
 M92           &  53  & 160  & 0.33 & 1.07  \\       
 M80           & 129  & 293  & 0.44 & 1.07  \\
 NGC~288       & 24   & 26   & 0.92 & 0.71  \\
\hline
\enddata
\tablecomments{$\dag$ In 47~Tuc the selection of BSS and HB has
been performed using the CMD plane $m_{{\rm F218W}},m_{{\rm
    F218W}}-m_{{\rm F439W}}$
instead of the $m_{{\rm F255W}},m_{{\rm F255W}}-m_{{\rm F336W}}$ plane.}
\tablecomments{$\ast$ Post core collapse cluster.}
\end{deluxetable}

\newdimen\minuswidth    
\setbox0=\hbox{$-$}
\begin{deluxetable}{ccccc}
\scriptsize \tablewidth{9cm}  
\tablecaption{\label{bb}
Expected number of single star ($N_{ss}$)
and binary-binary ($N_{bb}$) encounters per Gyr}
\startdata 
\hline
\hline
                                 \\
Cluster  & $a_{hs}(AU)$ & $\sigma_0$&$N_{ss}$  & $N_{bb}$ \\							
 & &$[km~s^{-1}]$ & & 	\\
\hline 
  M13       & 0.39 &7.1 & 0.07    &  0.17  \\
  NGC~6266  & 0.10 &14.3&   3.19   &   2.07 \\
  NGC~104   & 0.15 &11.5& 	1 & 	  1 \\
  NGC~6752  & 0.98 & 4.5&  0.16   &  1.04 \\
  M10       & 0.45 &6.6 & 0.05    & 0.11 \\
   M3       & 0.63 &5.6 &  0.16   &  0.70 \\
  M92       & 0.57 &5.9 &  0.28   &  1.04    \\
  M80       & 0.13 &12.4&  0.33   &  0.29 \\
 NGC~288    & 2.36 &2.9 & 	0 & 0.02   \\
\hline
\enddata
\end{deluxetable}

\newdimen\minuswidth    
\setbox0=\hbox{$-$}
\begin{deluxetable}{cccccc}
\scriptsize \tablewidth{12.5cm}  
\tablecaption{\label{coll}
MSP content, destruction and formation rate}
\startdata 
\hline
\hline
                                 \\
 & Isolated  & Binary & & & \\
Cluster  & PSRs  & PSRs & ${\cal R}_{\rm form}$ & ${\cal R}_{\rm dist}$ & ${\cal R}_{\rm form}/{\cal R}_{\rm dist}$ \\		
\\
\hline 
 M13     & 2  & 3  & $6.51\times10^{-2}$  &   $7.18\times10^{-2}$   &   $1.1$ \\
 NGC~6266 & 0  & 6  & $2.65$              &   $1.09$               &   $2.97$ \\
 NGC~104  & 7  & 15 & $1.00$              &   $1.00$ &   $1.00$ \\
 NGC~6752 & 4  & 1  & $0.18$              &   $6.72$ &   $3.21\times10^{-2}$ \\
 M10     & 0  & 0  & $4.61\times10^{-2}$  &   $0.14$               &   $0.40$ \\
 M3      & 0  & 4$\dag$ & $0.11$         &   $8.83\times10^{-2}$   &   $1.58$ \\
 M92     & 0  & 0  & $0.21$              &   $0.47$               &   $0.56$ \\
 M80     & 0  & 0  & $0.51$              &   $1.97$               &   $0.32$ \\
 NGC288  & 0  & 0  & $1.6\times10^{-3}$   &   $4.72\times10^{-3}$   &   $0.41$ \\
\hline
\enddata
\tablecomments{$\dag$ For 3 of these pulsars no timing/orbital
solution has been published yet.}
\end{deluxetable}

\newpage

\newpage

\newpage

\newpage

\newpage

\newpage

\end{document}